\documentclass[twocolumn,amsmath]{revtex4}
\usepackage{graphicx}

\begin{document}

\title{Symmetry breaking due to Dzyaloshinsky-Moriya interactions in the kagom\'e lattice}
\author{M.~Elhajal}
\email{elhajal@polycnrs-gre.fr}
\author{B.~Canals}
\email{canals@polycnrs-gre.fr}
\author{C.~Lacroix}
\email{lacroix@polycnrs-gre.fr}
\affiliation{Laboratoire Louis N\'eel, CNRS, BP 166, 38042 Grenoble Cedex 9, France}
\date{12 February 2002}

\begin{abstract}
Due to the particular geometry of the kagom\'e lattice, it is shown that antisymmetric Dzyaloshinsky-Moriya interactions are allowed and induce magnetic ordering.
The symmetry of the obtained low temperature magnetic phases are studied through mean field approximation and classical Mont\'e Carlo simulations.
A phase diagram relating the geometry of the interaction and the ordering temperature has been derived.
The order of magnitude of the anisotropies due to Dzyaloshinsky-Moriya interactions are more important than in non-frustrated magnets, which enhances its appearance 
in real systems. 
Application to the jarosites compounds is proposed.
In particular, the low temperature behaviors of the Fe and Cr-based jarosites are correctly described by this model.
\end{abstract}

\maketitle

\section{Introduction}

In the last few years, geometrically frustrated antiferromagnets have been the subject of much experimental and theoretical works \cite{diep94}.
Up to now, the most extensively studied are the kagom\'e ($D=2$) and the pyrochlore ($D=3$) antiferromagnets.
Both are expected to have disordered classical and quantum ground states and behave as spin liquids \cite{Chalker,Lecheminant}.
In such cases, it is expected that any small perturbation may have a strong effect on the ground sate manifold.
These perturbations can arise from thermal or quantum fluctuations, anisotropy, longer range interactions, etc.
This paper focus on the kagom\'e lattice where it is shown that Dzyaloshinsky-Moriya interactions (DMI) may be present.
As a consequence, magnetic ordering can occur at low temperature.
In particular the magnetic structures of the Fe and Cr jarosites \cite{Wills, Inami2, Wills2, Inami} are explored and it is proposed that DMI can explain the low 
temperature behaviors of these compounds.

DMI may be present in magnetic systems if there is no inversion center between two magnetic sites.
Such interactions\ between two sites i and j are defined by a vector $\mathbf{D}_{ij}$: $H_{ij}$ = $\mathbf{D}_{ij}.(\mathbf{S}_i\times\mathbf{S}_j)$.
It was Moriya \cite{Moriya} that clarified the conditions for the existence of these interactions and he gave some rules determining the possible directions of 
$\mathbf{D}_{ij}$.
He proposed a microscopic derivation of these interactions based on Anderson's formalism of superexchange, including spin-orbit coupling.
Other mecanisms were proposed for metallic systems with RKKY exchange interactions \cite{Smith,Fert}.

In fact the most general spin hamiltonian for two neighboring spin-1/2 magnetic ions is given by:
\begin{equation}
\label{hamiltonien_entier}
H_{ij}=J_{ij}\mathbf{S}_i.\mathbf{S}_j+\mathbf{D}_{ij}.(\mathbf{S}_i\times\mathbf{S}_j)+\mathbf{S}_i\overleftrightarrow{A_{ij}}\mathbf{S}_j
\end{equation}
where the second term is the antisymmetric DMI and the last term $\overleftrightarrow{A_{ij}}$ is an anisotropic symmetric exchange interaction.
Only antiferromagnetic isotropic exchange (J$_{ij}$\textgreater 0) will be considered in the following.

In antiferromagnetic oxydes such as $\alpha$-Fe$_2$O$_3$ \cite{Dzyaloshinsky}, La$_2$CuO$_4$ \cite{Coffey}, DMI is responsible for weak ferromagnetism.
In some Fe and Cr jarosites, a small ferromagnetic component was observed even when the main exchange interaction is antiferromagnetic ($\theta_{CW}\sim$-700~K for 
Fe jarosites). The low temperature magnetic structure is a long range ordered state where all the spins have the same component in the direction perpendicular to the 
kagom\'e plane, giving rise to weak ferromagnetism. The in-plane components of the spins form a {\bf q}={\bf 0} structure, the three spins of the triangular 
magnetic unit cell being at 120\textsuperscript o one from another. Only one of the two possible chiralities is observed. Depending on the jarosite, the out-of-plane 
component may vanish \cite{Wills}. 

\label{structure_jarosites}
Each of the magnetic (Fe and Cr) atoms which form the kagom\'e lattice is surrounded by an octahedron of oxygen atoms, and two neighbouring octahedra share one 
oxygen atom which mediates the superexchange interaction between the magnetic sites. 
These octahedra are responsible for the crystalline electrical field on the magnetic atoms. 
They are slightly distorted and their local axial axes are tilted with respect to the normal of the kagom\'e plane.

In section \ref{section:general}, some aspects of DMI specific to the kagom\'e lattice are discussed and a microscopic derivation of the DMI is made. Sections 
\ref{section:D_perp} and \ref{section:D_incl} deal with the magnetic properties due to the two types of possible DMI. The results are compared to the magnetic 
structures of the jarosites in section \ref{ref:experience}.

\section{\label{section:general} DM interactions in the kagom\'e lattice}

As a consequence of the hamiltonian being invariant under the symmetry operations which leave the lattice invariant, the direction of $\mathbf{D}_{ij}$ is 
geometrically constrained and follows rules explicited by Moriya \cite{Moriya}.
In the next section, Moriya's rules are applied to the kagom\'e lattice and to the related jarosites.
In section \ref{subsection:remarks}, an estimation of the order of magnitude of the different terms in (\ref{hamiltonien_entier}) is made. 
Section \ref{subsection:derivation} gives a microscopic derivation of DMI following Moriya's formalism and taking into account the peculiar structure and the 
environment of magnetic atoms in the jarosites.

\subsection{Application of symmetry rules} 

Two of Moriya's rules give useful informations about the $\mathbf{D}_{ij}$ in the kagom\'e lattice. First, the middle point between two sites is not a center of 
inversion for the kagom\'e lattice, so DMI are not forbidden by the symmetry of the lattice.
Furthermore, in a perfect kagom\'e lattice, the $\mathbf{D}_{ij}$ can only be perpendicular to the kagom\'e plane since this plane is a mirror plane. 
These symmetry considerations determine the axis of all the $\mathbf{D}_{ij}$ vectors (if DMI exist), but not their directions nor their values, which will depend 
on microscopic details.

In the jarosites, the symmetry is lowered because the octahedra of oxygen atoms which surround the magnetic sites are tilted \cite{Wills2, Inami}. 
The local axial axes of these tetrahedra are not exactly perpendicular to the kagom\'{e} plane, and the kagom\'{e} plane is then no longer a mirror plane for the 
lattice when the non-magnetic atoms are considered. 
These non-magnetic oxygen atoms that make up the coordination octahedra must be taken into account because they are responsible for the crystalline electric field 
on the magnetic atoms, and are involved in the superexchange mechanism between these magnetic sites. 
Applying Moriya's rules to the jarosites crystal constrains the $\mathbf{D}_{ij}$ vectors to be in the plane perpendicular to the bond (ij) since this is a mirror 
plane of the jarosite structure.

In both the pure kagom\'{e} and jarosite structures, the $\mathbf{D}_{ij}$ are not completely determined by symmetry. However, if we fix arbitrarily one of the 
$\mathbf{D}_{ij}$, then all the others are fixed by the three-fold rotation axis perpendicular to the kagom\'{e} plane that passes through the center of the 
triangles of the kagom\'{e} lattice.

Moriya's rules constraint the $\mathbf{D}_{ij}$ vectors with the help of the symmetry of the lattice, but they are not a proof of the existence of DMI in the 
kagom\'e lattice (or the related jarosites), they just express the fact that \emph{if} DMI are present, then the $\mathbf{D}_{ij}$ will necessarily be 
restricted to some set of possible vectors.

\subsection{\label{subsection:remarks}General considerations}

Taking into account the superexchange mecanism, the isotropic exchange $J_{ij}$ is proportional to $\frac{t_{ij}^{2}}{U}$ ($t_{ij}$ being the intersite hopping and 
$U$ the on-site Coulomb repulsion), while it was shown by Moriya that $|\mathbf{D}_{ij}|$ is proportional to $\frac{\lambda t_{ij}^{2}}{\Delta U}$ ($\lambda $ being 
the spin-orbit coupling and $\Delta$ the crystal field splitting) and $\overleftrightarrow{A}_{ij}$ is proportional to $\frac{\lambda ^{2}t_{ij}^{2}}{\Delta^2 U}$.
This last term, being one order of magnitude smaller than the DMI is often neglected.
However, it was suggested by Shekhtman \emph{et al.} \cite{Shekhtman} that it plays an important role, as does the DMI, because they are responsible for 
anisotropies respectively proportionnal to $\frac{D^2}{J}$ and $A$ which are of the same order of magnitude.
This argument is based on the assumption that the isotropic exchange $J$, which is the dominant interaction favours collinear configurations, and is untrue in the 
case of the kagom\'e lattice due to its frustration.
In a collinear structure DMI are in competition with isotropic exchange $J$ resulting in a $\frac{D^2}{J}$ anisotropy, whereas anisotropic symmetrical exchange 
defines easy axes or planes but is not contradictory with a collinear structure, resulting in a $\sim$A anisotropy.
Considering different possible $\overleftrightarrow{A}_{ij}$, we find that these arguments generally do not hold in the case of a non-collinear structure such as 
those found on the kagom\'e lattice.
Rather, the anisotropies are respectively of the order of $D$ and $A$, which is the reason why only DMI will be taken into account in this work.
The anisotropic exchange $\overleftrightarrow{A}$ is considered to enter the problem at the next order of perturbation theory.

Shekhtman \emph{et al.} also showed that under specific symmetry conditions it was possible to map the total spin hamiltonian of Eq. (\ref{hamiltonien_entier}) onto 
an isotropic Heisenberg hamiltonian (only first term of Eq. (\ref{hamiltonien_entier})) thus recovering a hidden rotational symmetry. 
These conditions are not fulfilled in the kagom\'e lattice and this mapping is not possible. 

\subsection{\label{subsection:derivation}Microscopic derivation of DMI using Moriya's technique}

\begin{figure}[tbp]
\includegraphics[width=7cm]{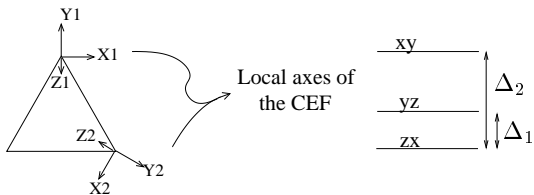}
\caption{Local axes of the crystalline electrical field considered in the microscopic derivation of the DMI. The axes are tilted by an angle $\alpha$ around the x 
axis toward the center of the triangle.}
\label{fig:champ_crist}
\end{figure}

As mentionned above, while application of Moriya's rules restricts the {\bf D} to some set of possible vectors, they are not proof of the existence of DMI.
In this section we derive DMI using the method proposed by Moriya assuming some microscopic situation (orbitals and crystalline electric field) which is arbitrary, 
but related to the case of the jarosites.
The method proposed by Moriya applies for localized magnetic electrons (insulators).
In order for the derivation to be manageable, we assume that the ground state is not degenerate due to cristal electrical field (except Kramers spin degeneracy) and 
each magnetic site is occupied by one electron.
In our case, we consider the t$_{2g}$ orbitals, whose degeneracy has been further lifted by some distortion of the octahedral environment.
The e$_g$ orbitals are assumed to be at much higher energy level (or completely filled and at much lower energy level).
The crystalline field scheme for the t$_{2g}$ is represented on Fig.~\ref{fig:champ_crist}.
The three states of each triangle are obtained from one another by a rotation of $\frac{2\pi}{3}$ around the axis perpendicular to the kagom\'e plane.

The octahedral symmetry and the 3d orbitals are relevant for the jarosites (see \ref{structure_jarosites}), and the lifting of degeneracy of the t$_{2g}$ orbitals 
is supported by the fact that the oxygen octahedra surrounding magnetic sites are distorted in the jarosites (see figures 2 and 3 of ref. \onlinecite{willsHFM}). 
The local axes of the cristalline field on each site have also been tilted towards the center of the triangles of the kagom\'e lattice by an angle $\alpha$, in order 
to fit with the jarosites symmetry.
Spin-orbit coupling is taken into account at first order in perturbation theory as it is expected to be much smaller than the crystalline electric field.
Next, the on-site Coulomb repulsion is assumed to be much higher than the hopping term, and the intersite hopping is introduced to second order in perturbation.
Doing so, we arrive at the following expression for the {\bf D}$_{12}$ vector 
\begin{eqnarray}
D_X & = & \sqrt3D_Y \\
D_Y & = & \frac{\lambda\sqrt3}{256U}\sin(2\alpha)f_1(\alpha) \nonumber \\
    &   & \times \left(\frac{f_2(\alpha)}{\Delta_1}+\frac{9}{\Delta_2}\left((dd\delta)-(dd\sigma)\right)\right) \\
D_Z & = & -\frac{\lambda\sqrt3\cos^2\alpha}{32U\Delta_1}f_1(\alpha)f_2(\alpha) 
\end{eqnarray}
where
\begin{eqnarray}
f_1(\alpha) & = & 4(dd\pi)-9(dd\sigma)\sin^2\alpha  \\
            &   & -3(dd\delta)(1+3\cos^2\alpha) \nonumber \\
f_2(\alpha) & = & 4(dd\pi)(3\cos^2\alpha-2)-9(dd\sigma)\sin^2\alpha  \\
            &   & +(dd\delta)(1+3\cos^2\alpha) \nonumber
\end{eqnarray}

$(dd\sigma)$, $(dd\pi)$ and $(dd\delta)$ are the transfer integrals as defined by Slater and Koster \cite{Slater}.

The obtained $\mathbf{D}_{ij}$ vectors are located in the plane perpendicular to the 
(ij) bond and are perpendicular to the kagom\'e plane if the octahedra are not tilted 
($\alpha=0$) as they should according to Moriya's rules.

\section{\label{section:D_perp}Case of \textbf{D} $\perp$ kagom\'e plane}

In this section the consequences of the DMI on the low temperature magnetic structure are explored in the case where \textbf{D} is perpendicular to the kagom\'e 
plane. Both cases D$_z$\textgreater0 and D$_z$\textless0 (z is the axis perpendicular to the kagom\'e plane and the convention is taken that the spins in the cross 
products appear rotating anticlockwise around the hexagons) are considered through mean field approximation and Mont\'{e} Carlo simulations with classical 
Heisenberg spins. 

\begin{figure}[tbp]
\includegraphics[width=7cm]{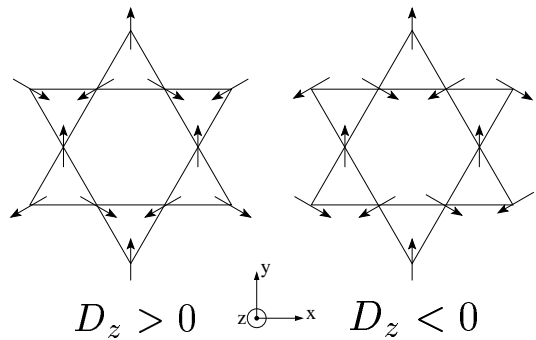}
\caption{$D\perp$ kagom\'e plane. 
The spins lie in the kagom\'e plane and the sign of D$_z$ selects the chirality. 
There is a global rotational degree of freedom around the z axis.}
\label{fig:chiralites}
\end{figure}

Looking for \textbf{q}=\textbf{0} structures, a restriction which will be justified by Mont\'{e} Carlo simulations, it is found that within mean field 
approximation, the system undergoes a phase transition to a long range ordered state. 
In the ordered state, all the spins lie in the kagom\'{e} plane, so one effect of the DMI is to act as an easy-plane anisotropy (as long as the first excitations 
are not taken into account e.g. looking only at the structure of the ground state).
As it can be easily seen by expanding the cross product with only a z component for the $\mathbf{D}_{ij}$, the hamiltonian is invariant under a global rotation of 
all the spins around the z axis. 
This degree of freedom is of course still present for the ground state. 
Since it is a \textbf{q}=\textbf{0} structure, all triangles have the same magnetic stucture. 
Depending on the sign of D$_{z}$, two chiralities are found and represented on Fig.~\ref{fig:chiralites}.

\begin{figure}[tbp]
\includegraphics[width=7cm]{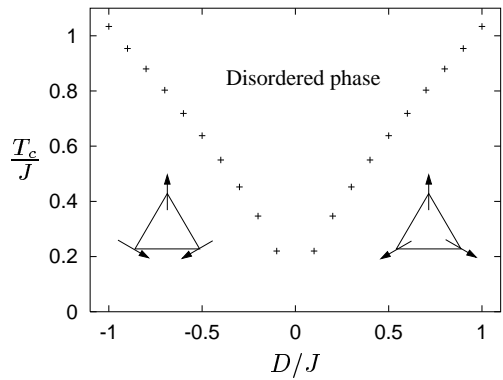}
\caption{$D\perp$ kagom\'e plane. The critical temperature is $\sim$ D.}
\label{fig:phases}
\end{figure}

In order to study the behaviour of this system at finite temperatures, Mont\'e Carlo simulations have been performed on finite size clusters with classical 
Heisenberg spins.
In particular, the behavior of the critical temperature as a function of $\frac{D}{J}$ is plotted in Fig.~\ref{fig:phases}. 
The critical temperature was taken as the maximum of the specific heat.

The  line  D$_{z}$=0 on Fig.~\ref{fig:phases} represents the well studied (classical) kagom\'{e} lattice with antiferromagnetic nearest neighbor exchange 
interactions. 
This system shows partial order: the coplanar states are assymptotically selected as temperature is lowered \cite{Chalker} (order by disorder). 
However, as soon as D has a finite value, even much smaller than J, a phase transition occurs at \emph{finite} temperature.
For this reason the low temperature magnetic structure will be governed by DMI. It must be emphasized that this argument would not be correct if DMI did not select a 
peculiar set of magnetic states, e.g. if the induced energy scale was renormalized to zero as it is the case for the exchange for the sole isotropic Heisenberg model.
This is supported by the fact that the critical temperature is almost linear in D, and only weakly dependent on the strength of the antiferromagnetic exchange 
interactions (J). The physical reason for this is that the long range ordered states which are imposed by DMI are part of the degenerate
ground states of the system with only isotropic exchange (J) interactions, even after a partial lifting of degeneracy due to thermal fluctuations. Thus, DMI has 
here a first order effect on a degenerate ground state, while usually DMI acts as a small perturbation on antiferromagnetically ordered ground state, leading to 
second order corrections. 

\section{\label{section:D_incl}Case of tilted oxygen tetrahedra}

In this section we study the general case where \textbf{D}$_{ij}$ is in the plane perpendicular to the (ij) bond. 
In order to clarify the discussion, we start by looking at the sole effect of the in-plane component of \textbf{D} before turning to the general case. 

\subsection{Case of D in the kagom\'e plane~: canted structure}

\begin{figure}[tbp]
\includegraphics[width=7cm]{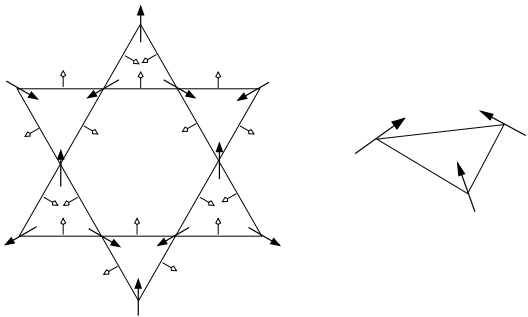}
\caption{In-plane \textbf{D} represented in the middle point between the magnetic sites and the associated magnetic structure. 
All the spins have a weak z component resulting in weak ferromagnetism. 
There is no global rotational degeneracy.}
\label{fig:dincline}
\end{figure}

The \textbf{D} vectors as well as the ground state magnetic configuration obtained by mean field approximation and Mont\'e Carlo simulations are represented on 
Fig.~\ref{fig:dincline}.

The structure has some similarities with the D$_z$\textgreater 0 case for \textbf{D}$\perp$ kagom\'e plane, and indeed it has the same chirality, but there are also
big differences. 
The spins do not lie in the kagom\'{e} plane anymore, but they all have the same out-of-plane z component, giving rise to weak ferromagnetism, perpendicular to the 
kagom\'{e} plane. 
Changing the sign of the in-plane component of \textbf{D} changes the sign of the z component of the spins as can be easily seen by expanding the cross product in 
the hamiltonian.
A second difference is that there is no longer a global rotational degree of freedom  and each spin points towards a fixed direction. 
In this case, DMI seems more to act like an easy-axis anisotropy. 
The ground state for the DMI alone does not belong to the ground state manifold  for the exchange interactions, since the sum of the spins on each triangle is not 
zero and there is now a competition between exchange and DMI. 
The consequence of this competition is that the angle between the spins and the kagom\'{e} plane depends now on $\frac{D}{J}$~:
this angle, as well as the weak ferromagnetic component, increases with $\frac{D}{J}$. 

\subsection{general case}

We now turn to the general case where \textbf{D}$_{ij}$ lies in the plane perpendicular to the (ij) bond. If D$_z$\textgreater 0 then the chirality which is
selected by D$_z$ is the same as the one which comes from the in-plane component D$_p$. Starting from the configuration for D$_z$=0,  the
effect of D$_z$ on the lowest energy configuration will be very much the same as J:  it will decrease the value of the angle between the spins and
the kagom\'{e} plane. The case D$_z$\textless 0 is more complicated~: the chirality favoured by D$_z$ is not the same as the one which is selected if only
the in-plane component of D is considered. The result is a competition between the in-plane and the out-of-plane components of \textbf{D}. When \textbf{D}
is almost in the kagom\'{e} plane with a small negative component along the z axis, then the canted structure is selected with a ferromagnetic moment
which increases with increase of the z component. There is a critical value of $\frac{D_{z}}{D_{p}}$, and if D$_{z}$ is negative enough, then the ground
state is no longer the canted structure, but the planar structure on the right of Fig.~\ref{fig:chiralites}, as in the case of  D $\perp $ kagom\'{e}
plane. This critical value depends on the strength of the exchange interactions J, since the latter favour the planar structure.

\begin{figure}[tbp]
\includegraphics[width=7cm]{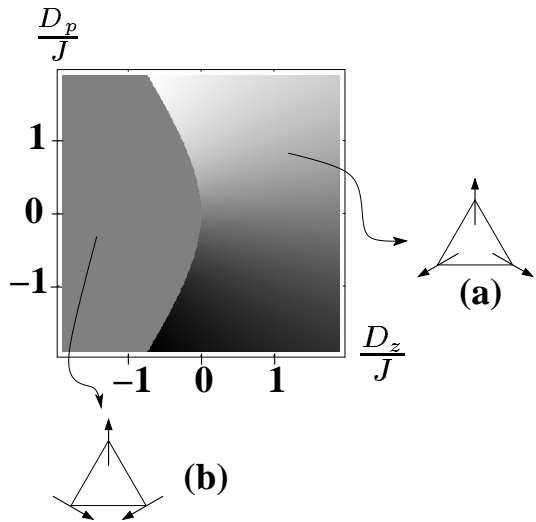}
\caption{Ground states obtained for different values of J (isotropic exchange), D$_p$ and D$_z$ (in plane and out of plane components of \textbf{D}). 
The intensity of the grey color represents $\eta$, the angle between the spins and the kagom\'e plane. The grey on the left part corresponds to $\eta=0$ (spins in 
the kagom\'e plane) while on the right part, the white corresponds to $\eta$\textgreater 0 and the black to $\eta$\textless 0.}
\label{fig:etatfond}
\end{figure}

It appears that the magnetic structures for \textbf{D} and -\textbf{D} are not always similar as in the case of \textbf{D} $\perp$ to the kagom\'e plane. 
They can not be deduced easily from each other.
This can also be seen on the hamiltonian~: when replacing \textbf{D} by -\textbf{D}, there is no simple transformation of the \textbf{S}$_i$ which would leave the 
hamiltonian invariant. 
This comes from the fact that the lattice is not bipartite, since in a bipartite lattice, exchanging the two sublattices would transform $\mathbf S_i\times\mathbf S_j$ into 
$-\mathbf S_i\times\mathbf S_j$ and leave the hamiltonian invariant. 
However, D$_p \to$~-D$_p$ corresponds to a trivial symmetry~: changing simultaneously S$_z \to$~-S$_z$ will leave the hamiltonian invariant, and so the non-trivial 
part of the transformation {\bf D}$\to$-{\bf D} comes from the z component D$_z$.
The discrepancy between the structures for {\bf D} and -{\bf D} is illustrated on Fig.~\ref{fig:etatfond}.
It represents the different ground states obtained for different values of the parameters D$_p$, D$_z$ and J. 

This figure was obtained by comparing the energies of the two possible chiralities (a) and (b), (see figure \ref{fig:etatfond}) on one triangle, and minimizing with 
respect to $\eta$, the angle between the spins and the kagom\'e plane and the angle $\varphi$ between the projection of the spins on the kagom\'e plane and their 
postion as represented on figure \ref{fig:etatfond}. 
The energies of the two triangles (a) and (b) are: 

\begin{equation}
\frac{E_a}{N}=\frac{J}{2}\left(1-3\cos\left(2\eta\right)\right)-\sqrt3 D_z\cos^2\eta-\sqrt3 D_p\sin(2\eta)\cos\varphi
\end{equation}

\begin{equation}
\frac{E_b}{N}=\frac{J}{2}\left(1-3\cos\left(2\eta\right)\right)+\sqrt3 D_z\cos^2\eta
\end{equation}

Several aspects already mentionned appear on these expressions. 
First, the sign of D$_z$ selects the chirality since it appears with a different sign in E$_a$ and E$_b$. 
D$_p$ appears to favour chirality (a) and weak ferromagnetism perpendicular to the kagom\'e plane ($\eta\neq0$). 
For chirality (a), if D$_p$=0, the spins lie in the kagom\'e plane ($\eta=0$), and there is a global rotational degree of freedom since the energy would not depend 
on $\varphi$. 
If chirality (b) is selected, then $\eta=0$ always minimizes the energy (even for $D_p\neq0$) and so only in-plane structures appear with this chirality and the 
global rotationnal degree of freedom is always present. 
Furthermore, in this case the energy does not depend on D$_p$.

When chirality (a) is selected, and $D_p\neq0$, the spins have an out of plane component with the angle between the spins and the kagom\'e plane being:

\begin{equation}
\tan(2\eta)=\frac{2D_p}{\sqrt3 J+D_z}
\end{equation}

In this case, $\varphi=0$ (no global rotationnal degree of freedom).  

\begin{figure}[tbp]
\includegraphics[width=7cm]{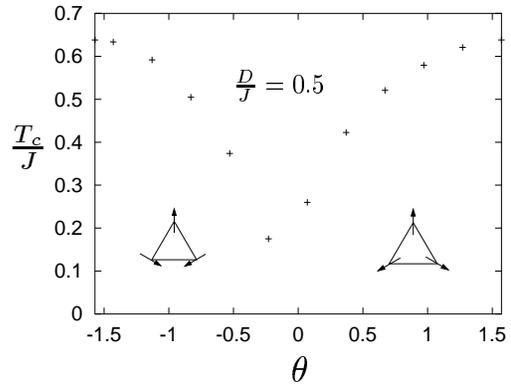}
\caption{Critical temperature as a function of the angle $\theta$ between {\bf D} and the kagom\'e plane, for a fixed value of D=$\sqrt{D^2_p+D^2_z}$}
\label{fig:theta}
\end{figure}

Figure \ref{fig:theta} shows the evolution of the critical temperature as a function of the angle $\theta$ between {\bf D} and the kagom\'e plane.

The critical temperature is lower when {\bf D} is in the kagom\'e plane ($\theta$=0) than when {\bf D} is perpendicular ($\theta = \pm \frac{\pi}{2}$). 
This is easily understood since for {\bf D} in the kagom\'e plane, DMI favours a canted structure, which is not a lowest energy configuration for the isotropic 
exchange part of the hamiltonian (J). For $\theta = \pm \frac{\pi}{2}$, DMI selects a coplanar structure, which also minimizes the isotropic exchange.  

The fact that the critical temperature is not symmetrical with respect to $\theta$=0 (Figure \ref{fig:theta}) also illustrates the different behaviour for 
\textbf{D} and -\textbf{D}. Indeed, changing $\theta\to -\theta$ corresponds to D$_z\to$ -D$_z$ which is the non-trivial part of the transformation 
\textbf{D}$\to$ -\textbf{D}, 
because D$_p\to$ -D$_p$ corresponds to the trivial symmetry S$_z \to$~-S$_z$. 
The fact that the critical temperature is always different for symmetrical values of $\theta$ with respect to 0 is due to the fact that D$_z$\textless 0 is in competition with the in-plane component of \textbf{D} since they tend to select different chiralities; on the contrary D$_z$\textgreater 0 and D$_p$ drive the system to 
the same chirality, hence a higher critical temperature is obtained for $\theta$\textgreater 0 than for $\theta$\textless 0. 

\section{\label{ref:experience}Magnetic structure of Fe and Cr jarosites}

We are interested here in the low temperature magnetic structure of the Fe and Cr based jarosites, where the magnetic ions (Fe$^{3+}$ and Cr$^{3+}$) form kagom\'e 
planes which are stacked one on another \cite{Wills, Inami, Wills2, Inami2, willsHFM} giving rise to a three dimensional structure. 
However, the magnetic behavior of the jarosites is believed to be essentially that of a (two dimensionnal) kagom\'e lattice, because the different kagom\'e planes 
are fairly well separated by several non-magnetic atoms implying that the super-exchange interactions between neighboring planes are much weaker than between 
magnetic sites belonging to the same kagom\'e plane.
These magnetic sites are surrounded by octahedra of oxygen atoms responsible for the crystalline electrical field and involved in the super-exchange interactions. 
These octahedra are slightly distorted and tilted (see figures 2 and 3 of \cite{willsHFM}) which is consistent with the microscopic electronic states we have 
considered for our microscopic derivation of DMI.

A long range ordered state is observed experimentally in these compounds \cite{Wills, Wills2, Inami, Inami2, willsHFM}, and the low temperature magnetic structure 
has an ordering wave vector \textbf k=(0,0,0) or \textbf k=(0,0,$\frac{3}{2}$), depending on the diamagnetic atoms present in the compound.
The third component of the \textbf k vector corresponds to the propagation of the magnetic structure from one kagom\'e plane to another and is 
due to weak inter-plane super-exchange interactions which were not considered in this article since they play no role in the ordering of one kagom\'e plane.

Henceforth the ordering of only one kagom\'e plane is considered.
The different observed magnetic structures have a \textbf q=\textbf 0 wave vector and always correspond to chirality (a).
Both planar and weak-ferromagnetic (all the spins having the same z-component along the axis perpendicular to the kagom\'e plane) structures are observed 
experimentally and both are obtained here theoretically.
The planar structure with chirality (b) obtained theoretically is not observed experimentally in these jarosites. 

The experimental structures can also be obtained introducing a single-ion anisotropy \cite{Inami, Inami2}, however this is not a relevant interaction at least 
in the case of Fe-jarosites. Indeed, the magnetic ions are Fe$^{3+}$, and so the 3d orbitals are half filled, with five electrons coupled giving rise to a spin 
S=$\frac{5}{2}$. Thus, the charge distribution is spherical for this orbital occupancy, and consequently prevents the 
appearance of a spin anisotropy. Also for Cr$^{3+}$ ions, the 3 t$_{2g}$ orbitals are filled (S=$\frac{3}{2}$) and the single ion anisotropy should be small.

\section{Conclusions}

It has been shown that DMI are allowed by the symmetry of the kagom\'e lattice and are relevant interactions whose effect is enhanced by the frustration of the 
predominant antiferromagnetic isotropic exchange. 
A microscopic derivation of these interactions for a schematic electronic configuration somewhat related to the jarosites structure was done. 
The magnetic properties due to DMI were studied through mean field approximation and Mont\'e Carlo simulations. 
Contrarily to the predominant antiferromagnetic isotropic exchange, 
DMI can induce several long range ordered magnetic structures depending on the different microscopic parameters J, D$_p$ and D$_z$. 
The low temperature magnetic structures of the Fe and Cr based jarosites can be explained by the presence of DMI. This would be the first example of compounds where 
magnetic ordering is not due to exchange interactions but to DMI.

\begin{acknowledgments}
It is a pleasure to acknowledge Dr Andrew Wills for fruitful discussions.
\end{acknowledgments}

\bibliographystyle{unsrt}

\end{document}